\documentclass[aps,prc,superscriptaddress,floats,floatfix,nofootinbib]{revtex4}
\pdfoutput=1

\usepackage{amsmath,amsfonts,amssymb,bm}
\usepackage{graphicx}
\usepackage{color}
\usepackage{multirow}
\usepackage{rotating}
\usepackage{float}
\usepackage{placeins}
\usepackage{adjustbox}
\usepackage{array}
\definecolor{purple}{rgb}{0.5,0,0.5}
\definecolor{blue}{rgb}{0.0,0,0.9}
\usepackage[colorlinks=true,
            pdfstartview=FitV,
            bookmarks=true,
            linkcolor=purple,
            citecolor=purple,
            urlcolor=blue,
            anchorcolor=black]{hyperref}

\newcommand{\OffShellMark}{\mkern0.4mu{}^{\mkern-1.0mu\raisebox{0.05ex}{\scalebox{0.52}[0.78]{\scriptsize\#}}}}
\newcommand{\ExpMark}{\mkern1mu{}^{\mathrm{E}}}

\begin{document}

\title{\Large Vector- and tensor-meson resonance contributions to $\chi_{cJ}\to PPV$ decays}
\author{Yuan-Guo Xu$^1$, Qin-Ze Song$^1$, Bo Lan$^{2,\dag}$, and Ru-Min Wang$^{1}$\\
	{\scriptsize $^1$School of Physics, Jiangxi Normal University, Nanchang, Jiangxi 330022, China}\\
	{\scriptsize $^2$College of Physics and Electronic Information, Nanchang Normal University, Nanchang, Jiangxi 330032, China}\\
	$^\dag${\scriptsize Corresponding author.~~Email:lanbo@ncnu.edu.cn}
}

\begin{abstract}
Light vector- and tensor-meson resonance contributions to the three-body decays $\chi_{cJ}\to P_1P_2V$ are studied in a quasi-two-body framework combined with SU(3) flavor relations for the two-body subprocesses, covering both the $\chi_{c0,1,2}\to VV\to P_1P_2V$ and $\chi_{c1,2}\to TP\to P_1P_2V$. Both the narrow-width approximation (NWA) and the finite-width effects (FWE) treatment are considered, the latter being essential for tensor subdecays that are kinematically closed at the nominal mass and contribute only through the high-mass tail of the resonance. The predicted resonance bands span several orders of magnitude, with the largest ones in the $\pi\pi\rho$, $\pi K\bar K^{*}$, and $K\bar K\phi$ final states, whereas the $K\bar K\omega$ and $\eta'KK^{*}$ channels are dominated by the tensor band and thus provide a clean window onto the $T\to PV$ tensor resonances. The band-by-band strengths and their ordering are intended as priors and channel priorities for future invariant-mass and amplitude analyses, and can be compared with the measured totals of several channels.

\end{abstract}

\maketitle

\section{Introduction}
\label{sec:introduction}

Charmonium can be used to study the strong interaction in the region where perturbative QCD is not sufficient. The spin-triplet $P$-wave states $\chi_{cJ}$ ($J=0,1,2$) have the same quark content and orbital angular momentum, but different total angular momenta. Their hadronic decays provide information on nonperturbative QCD and on the decay mechanisms of the $\chi_{cJ}$ states~\cite{Brambilla:2010cs,Voloshin:2007dx,Brambilla:2004wf,Mitchell:2012chi,Bodwin:1994jh,Vairo:2003gh}.

The $\chi_{cJ}$ states are mainly produced through the radiative transition $\psi(3686)\to\gamma\chi_{cJ}$. The data samples collected by BESIII in 2009, 2012, and 2021 amount to approximately $2.7\times10^{9}$ $\psi(3686)$ events in total~\cite{BESIII:2013Psi3686,BESIII:2018Psi3686,BESIII:2024Psi3686}. Detailed studies of exclusive $\chi_{cJ}$ hadronic decays and their intermediate resonance structures are therefore possible~\cite{BESIII:2020Future,BESIII:2010NeutralPairs,BESIII:2011VectorPairs,BESIII:2023PhiPhiHelicity}.

Experimentally, several exclusive $\chi_{cJ}$ decays with $P_1P_2V$ final states have been measured~\cite{BES:2006PhiKK,BESIII:2013OmegaKK}. Intermediate light-meson resonances have also been observed in multihadron charmonium decays~\cite{CLEO:2007ThreeBody,Belle:2008FourMeson,BESIII:2015PhiKstarK}. Invariant-mass distributions, Dalitz plots, and partial-wave analyses can be used to identify these intermediate resonances~\cite{BESIII:2023Eta1405,Nakamura:2024ThreeBody,CLEO:2011Amplitude,BESIII:2017Amplitude,BESIII:2014EtaPrimeKK,Liang:2016hmr}. However, a measured total branching ratio does not separate different resonance contributions. Experimental information on individual resonance contributions is still limited. A systematic study of these contributions is therefore useful for identifying important channels and guiding future measurements.

Three-body decays contain more dynamical information than two-body decays but are more difficult to describe theoretically. When a specific intermediate resonance dominates, the decay can be described in a quasi-two-body form as two sequential two-body processes~\cite{cc2PPP,Capstick:1993kb,Wang:2016rlo,Zhou:2025ThreeBodyD}. The resonance line shape and finite width can affect the resulting three-body decay rate, especially for broad resonances or near dynamical thresholds~\cite{Nakamura:2024ThreeBody,Cheng:2020iwk,Wang:2022vga}. Several decay processes and amplitudes relevant to this work have been studied in earlier publications. The two-body $\chi_{cJ}\to VV$ and $\chi_{cJ}\to TP$ decays were analyzed in Ref.~\cite{Lan:2025wkf}. The $T\to PV$ amplitudes were studied in Ref.~\cite{T2PV}, and this sector is updated here. Light-meson resonance contributions to $\chi_{c1,2}\to P_1P_2P_3$ were investigated in Ref.~\cite{cc2PPP}. Unlike the $P_1P_2P_3$ final states, which contain no vector meson, the $P_1P_2V$ modes provide direct access to the $T\to PV$ tensor resonances. In the tensor sector, the $T\to PV$ rate has a strong $p^5$ phase-space dependence, so that threshold and off-shell effects can be important. The main text is focused on the $P_1P_2V$ cascade processes and their phenomenology, while the previously established SU(3) ingredients and the detailed $T\to PV$ amplitude relations are collected in the Appendices.

In this work, the resonance decay processes $\chi_{c0,1,2}\to VV\to P_1P_2V$ and $\chi_{c1,2}\to TP\to P_1P_2V$ are studied. The corresponding resonance contributions are evaluated both in the narrow-width approximation and with finite-width effects included. For tensor subdecays that are kinematically closed at the nominal resonance mass, contributions from the kinematically allowed high-mass tail of the resonance are also estimated. Particular attention is paid to the relative sizes of the vector- and tensor-resonance contributions and to channels that can be tested in future amplitude analyses.

The $P_1P_2V$ resonance framework is given in Sec.~\ref{sec:framework}, where the narrow-width approximation and finite-width treatment are discussed together. The numerical results and the main phenomenological features are presented in Sec.~\ref{sec:phenomenology}. The updated $T\to PV$ amplitudes and inputs, the complete channel-by-channel results, and the SU(3) conventions are collected in the Appendices.

\section{Resonance framework for $\chi_{cJ}\to PPV$}
\label{sec:framework}

\subsection{Quasi-two-body description}

When a specific resonance contribution is considered, a quasi-two-body description is commonly used~\cite{Capstick:1993kb,Wang:2016rlo}. The three-body process is then factorized into two sequential two-body decays. For the $P_1P_2V$ final states considered here, the intermediate resonance is denoted by $R$. The meson $M_3$ produced in the first-step decay is treated as the spectator meson. The decay process is written as
\begin{equation}
\chi_{cJ}\to R M_3\to M_1M_2M_3,
\end{equation}
where $R=V$ or $T$. In the narrow-width approximation, the width of the intermediate resonance is assumed to be much smaller than its mass, $\Gamma_R\ll m_R$. If interference among different resonance amplitudes is neglected, the three-body branching ratio can be written as the product of the two two-body branching ratios~\cite{cc2PPP,Cheng:2020iwk},
\begin{equation}
\mathcal{B}(\chi_{cJ}\to R M_3\to M_1M_2M_3)
=
\mathcal{B}(\chi_{cJ}\to R M_3)
\times\mathcal{B}(R\to M_1M_2).
\label{eq:nwa}
\end{equation}

For the first-step two-body decays, the branching ratios of the $\chi_{cJ}\to VV$ and $\chi_{cJ}\to TP$ processes are given by~\cite{Lan:2025wkf}
\begin{equation}
\mathcal{B}(\chi_{cJ}\to M_1M_2)
=\frac{|\vec p|}{8\pi M_{\chi_{cJ}}^2\Gamma_{\chi_{cJ}}}
\left|\mathcal{A}(\chi_{cJ}\to M_1M_2)\right|^2,
\label{eq:two_body_chic}
\end{equation}
with
\begin{equation}
|\vec p|=\frac{\lambda^{1/2}(M_{\chi_{cJ}}^2,m_{M_1}^2,m_{M_2}^2)}{2M_{\chi_{cJ}}},
\end{equation}
and $\lambda(a,b,c)=a^2+b^2+c^2-2ab-2ac-2bc$.
For the subsequent resonance decays, the branching-ratio formulas for
$V\to P_1P_2$ and $T\to PV$ are written as~\cite{Cheng:2020iwk,Li:2015TensorDecay}
\begin{align}
\mathcal{B}(V\to P_1P_2)&=\frac{\tau_V {p^3_c}}{6\pi m_V^2}\,|g_{V\to P_1P_2}|^2,\label{eq:V2PP}\\
\mathcal{B}(T\to PV)&=\frac{\tau_T p_c'^5}{20\pi}\,|g_{T\to PV}|^2,\label{eq:T2PV}
\end{align}
where
\begin{equation}
p_c=\frac{\lambda^{1/2}(m_V^2,m_{P_1}^2,m_{P_2}^2)}{2m_V},
\qquad
p_c'=\frac{\lambda^{1/2}(m_T^2,m_P^2,m_V^2)}{2m_T}.
\end{equation}
Here, $\tau_R$ denotes the lifetime of the resonance $R$, and
$g_{R\to M_1M_2}$ is the strong coupling constant for the decay
$R\to M_1M_2$.

The narrow-width approximation is generally reliable when the decay
width of the resonance is much smaller than its mass. However, the change in phase space over the resonance mass distribution is not included. To account for finite-width effects, a relativistic Breit--Wigner line shape is used and an integration over the resonance mass distribution is performed. Following
Ref.~\cite{cc2PPP}, only the finite width of the
intermediate resonance is included. The branching ratio is
\begin{equation}
\mathcal{B}(\chi_{cJ}\to R M_3\to M_1M_2M_3)
	=
\frac{1}{\pi}
\int_{(m_R-n\Gamma_R)^2}^{(m_R+n\Gamma_R)^2}
\frac{
\mathcal{B}(\chi_{cJ}\to R M_3,\sqrt{t_R})\,
\mathcal{B}(R\to M_1M_2,\sqrt{t_R})\,
		m_R\Gamma_R
	}{
		(t_R-m_R^2)^2+m_R^2\Gamma_R^2
	}
\,dt_R.
\label{eq:fwe_general}
\end{equation}
Here, $t_R$ is the squared invariant mass of the intermediate resonance. 

Following the prescriptions adopted in
Refs.~\cite{T2PV,Wang:2022fbk}, $n=2$ is used throughout the numerical
analysis for decay channels with limited available phase space.

For some phase-space-limited $T\to PV$ channels, the decay is forbidden at the nominal tensor-meson mass. This occurs when $m_T<m_P+m_V$. The narrow-width approximation cannot be applied to these channels. To estimate the off-shell contribution from the finite tensor-meson width, the tensor-meson mass distribution is retained and the integration is performed over the allowed region of dynamics. The effective off-shell branching ratio for the subdecay $T\to PV$ is~\cite{Roca:2003uk}
\begin{equation}
\mathcal{B}^{\rm off}_{T\to PV}
	=-\frac{1}{N_T\pi}
\int_{(m_T-n\Gamma_T)^2}^{(m_T+n\Gamma_T)^2}
	ds_T\,
	{\rm Im}\left\{
\frac{1}{s_T-m_T^2+i m_T\Gamma_T}
\right\}
\mathcal{B}_{T\to PV}(\sqrt{s_T})
\Theta(\sqrt{s_T}-m_P-m_V).
\label{eq:off_shell_T}
\end{equation}
Here, $s_T$ is the integration variable for the tensor-meson mass squared, and $\Theta$ is the step function that imposes the dynamical threshold. The quantity $\mathcal{B}_{T\to PV}(\sqrt{s_T})$ is evaluated at the tensor-meson mass $\sqrt{s_T}$, and $N_T$ is the normalization factor for the tensor-meson mass distribution over the chosen integration interval.

\subsection{SU(3) framework and parameter inputs}

%The first-step branching ratios for $\chi_{cJ}\to VV$ and $\chi_{cJ}\to TP$ are taken from Ref.~\cite{Lan:2025wkf}. The $V\to P_1P_2$ amplitudes and the corresponding SU(3)-breaking treatment follow Ref.~\cite{cc2PPP}. For the $T\to PV$ sector, the updated amplitude relations and numerical constraints adopted in the present analysis are used~\cite{T2PV}.

Because these decays occur in the nonperturbative QCD region, their amplitudes are difficult to calculate with perturbative QCD. The SU(3) flavor symmetry approach is used to parameterize the amplitudes of different two-body decay channels~\cite{Haber:1985cv,Mo:2024jjd,Mo:2024zsa}. The amplitudes for the first-step $\chi_{cJ}\to VV/TP$ processes are parameterized as~\cite{Lan:2025wkf}
\begin{equation}
	\mathcal{A}(\chi_{cJ}\to M_1M_2)
	=
	a^M_{1J}M^i_{1j}M^j_{2i}
	+a^M_{2J}M^i_{1i}M^j_{2j}
	+b^M_J M^i_{1j}W^j_kM^k_{2i}.
	\label{eq:chi_su3}
\end{equation}
Here, $M_1M_2$ is either $VV$ or $TP$. The flavor indices $i,j,k=1,2,3$ correspond to the light-quark flavors $u$, $d$, and $s$, respectively. The coefficients $a^M_{1J}$ and $a^M_{2J}$ parameterize the SU(3)-symmetric contributions, whereas $b^M_J$ parameterizes SU(3) flavor breaking. The SU(3) breaking mainly comes from the mass difference between the strange quark and the nearly degenerate $u$ and $d$ quarks. The explicit pseudoscalar-, vector-, and tensor-meson matrices, together with the flavor-breaking matrix $W$, are given in the Appendix~\ref{app:su3details}.

For the resonance decay $V\to P_1P_2$, the strong coupling constants are written in the SU(3) flavor framework as
\begin{align}
	g^s_{V\to P_1P_2}=&\;g_{10}^{V}V^i_jP^i_kP^k_j
	+g_{20}^{V}V^i_jP^i_jP^k_k,\notag\\
	g^b_{V\to P_1P_2}=&\;g_{11}^{V}V^i_jW^i_mP^m_kP^k_j
	+g_{12}^{V}V^i_jP^i_kW^k_mP^m_j
	+g_{21}^{V}V^i_jW^i_mP^m_jP^k_k.
	\label{eq:VPP_coeff}
\end{align}
For the two-body $T\to PV$ decays, the strong coupling constants are
\begin{align}
	g^s_{T\to PV}=&\;g_{10}^{T}T^i_jM^i_k{M^\prime}^k_j
	+g_{20}^{T}T^i_jM^i_j{M^\prime}^k_k,\notag\\
	g^b_{T\to PV}=&\;g_{11}^{T}T^i_jW^i_mM^m_k{M^\prime}^k_j
	+g_{12}^{T}T^i_jM^i_kW^k_m{M^\prime}^m_j
	+g_{21}^{T}T^i_jW^i_mM^m_j{M^\prime}^k_k,
	\label{eq:TPV_coeff}
\end{align}
where $M,M^{\prime}=V$ or $P$. The superscripts $s$ and $b$ denote the SU(3)-symmetric and symmetry-breaking terms, respectively. The parameters $g_{10}$ and $g_{20}$ are SU(3)-symmetric, whereas $g_{11}$, $g_{12}$, and $g_{21}$ describe SU(3) flavor breaking. 

For $V\to P_1P_2$ processes, the available data are sufficient to keep both the symmetric and symmetry-breaking contributions; detailed results can be found in Ref.~\cite{cc2PPP}. For $T\to PV$ processes, the limited data give large uncertainties in the symmetry-breaking terms. We therefore keep only the symmetric contributions in the numerical analysis of the $T\to PV$ channels, and the updated amplitude relations and numerical constraints adopted in the present analysis are used. Since the basic SU(3) construction has been discussed previously~\cite{T2PV}, the detailed relations are not repeated in the main text. They are given in Appendix~\ref{app:t2pv} together with the updated branching ratio inputs.

\section{Phenomenology of the $PPV$ resonance contributions}
\label{sec:phenomenology}

For the decay processes $\chi_{cJ}\to VV\to P_1P_2V$ and $\chi_{cJ}\to TP\to P_1P_2V$, the branching ratios of the first-step two-body decays $\chi_{cJ}\to M_1M_2$ are taken from Ref.~\cite{Lan:2025wkf}. The currently available experimental values of the masses, lifetimes, and branching ratios are taken from the 2026 Particle Data Group (PDG)~\cite{PDG}. Input quantities with quoted uncertainties are varied within their corresponding $1\sigma$ error ranges, while upper-limit inputs are sampled within their allowed ranges. For each channel, at least 10,000 accepted parameter sets are generated by Monte Carlo sampling. The detailed $V\to PP$ inputs and the previously established SU(3) conventions are collected in Appendix~\ref{app:su3details}, and the experimental data and the corresponding theoretical predictions for the $T\to PV$ decays are summarized in Appendix~\ref{app:t2pv}.

 The modes $K_2^*(1430)\to K^*\eta$, $K_2^*(1430)\to\phi K$, $a_2(1320)\to K^*K$, and $f_2(1270)\to K^*K$ are kinematically closed at the nominal tensor mass. Their contributions to the three-body decays are obtained only through the finite-width treatment in Eq.~(\ref{eq:off_shell_T}). In addition, the measured branching ratios of $a_{2}(1320) \to \rho\pi$, $K^{*}_{2}(1430) \to \rho K$, $K^{*}_{2}(1430)\to  K^*\pi $, and $K^{*}_{2}(1430) \to \omega K$ are used separately to constrain the parameter combinations relevant to the corresponding groups of decay modes. The other $T \to PV$ branching ratios are predicted from the accepted parameter sets using the amplitude relations collected in Appendix~\ref{app:t2pv}.

Tables~\ref{Tab:B2PVlv1}--\ref{Tab:Brchi22PPVpartII} list the results for the three-body decays $\chi_{cJ}\to P_1P_2V$ under the NWA and with FWE. Among the $\chi_{c0}$ channels in Table~\ref{Tab:B2PVlv1}, only $\mathcal{B}(\chi_{c0}\to\pi^-K^+\bar K^{*0})=(7.0\pm1.5)\times10^{-3}$ has so far been measured~\cite{PDG}. The available experimental data for the $\chi_{c1,2}$ decays are also included in the corresponding tables. For each pair of charge-conjugate contributions with charged intermediate resonances, we list only the contribution from the negatively charged resonance. The positive-charge contribution is related by charge conjugation and is not listed. Contributions from neutral intermediate resonances are listed separately.

\begin{table}[H]
	\centering
	\renewcommand\arraystretch{1.00}
	\setlength{\tabcolsep}{2.6pt}
	\caption{Branching ratios for the $\chi_{c0}\to P_1P_2V$ decays arising from vector-meson resonance contributions. NWA denotes results under the narrow-width approximation, whereas FWE denotes results with finite-width effects included.}
	\begingroup
	\newcommand{\XCModeCol}[1]{\makebox[0.190\textwidth][l]{#1}}
	\newcommand{\XCResultCol}[1]{\makebox[0.380\textwidth][c]{#1}}
	\begin{tabular}{l|l|l}
		\hline\hline
		\XCModeCol{Decay mode} & \XCResultCol{$V$-resonance contribution (NWA)} & \XCResultCol{$V$-resonance contribution (FWE)} \\
		\hline
		\XCModeCol{$\chi_{c0}\to \eta\pi^-\rho^+$} & \XCResultCol{$<2.30\times10^{-5}_{[\rho^-\to\pi^-\eta]}$} & \XCResultCol{$<2.84\times10^{-5}_{[\rho^-\to\pi^-\eta]}$} \\
		\hline
		\XCModeCol{$\chi_{c0}\to \pi^0\pi^-\rho^+$} & \XCResultCol{$(2.28\pm1.56)\times10^{-3}_{[\rho^-\to\pi^-\pi^0]}$} & \XCResultCol{$(1.92\pm1.32)\times10^{-3}_{[\rho^-\to\pi^-\pi^0]}$} \\
		\hline
		\XCModeCol{$\chi_{c0}\to \eta\pi^0\rho^0$} & \XCResultCol{$<1.14\times10^{-5}_{[\rho^0\to\pi^0\eta]}$} & \XCResultCol{$<1.41\times10^{-5}_{[\rho^0\to\pi^0\eta]}$} \\
		\hline
		\XCModeCol{$\chi_{c0}\to \pi^-\pi^+\rho^0$} & \XCResultCol{$(1.13\pm0.77)\times10^{-3}_{[\rho^0\to\pi^+\pi^-]}$} & \XCResultCol{$(9.44\pm6.46)\times10^{-4}_{[\rho^0\to\pi^+\pi^-]}$} \\
		\hline
		\XCModeCol{$\chi_{c0}\to \pi^0K^-K^{*+}$} & \XCResultCol{$(5.66\pm2.00)\times10^{-4}_{[K^{*-}\to\pi^0K^-]}$} & \XCResultCol{$(4.86\pm1.88)\times10^{-4}_{[K^{*-}\to\pi^0K^-]}$} \\
		\hline
		\XCModeCol{$\chi_{c0}\to \pi^-\bar{K}^{0}K^{*+}$} & \XCResultCol{$(1.13\pm0.40)\times10^{-3}_{[K^{*-}\to \bar{K}^{0}\pi^-]}$} & \XCResultCol{$(9.73\pm3.76)\times10^{-4}_{[K^{*-}\to \bar{K}^{0}\pi^-]}$} \\
		\hline
		\XCModeCol{$\chi_{c0}\to \pi^-K^+\bar{K}^{*0}$} & \XCResultCol{$(1.13\pm0.40)\times10^{-3}_{[K^{*0}\to\pi^-K^+]}$} & \XCResultCol{$(9.56\pm3.37)\times10^{-4}_{[K^{*0}\to\pi^-K^+]}$} \\
		\hline
		\XCModeCol{$\chi_{c0}\to \pi^+K^-K^{*0}$} & \XCResultCol{$(1.13\pm0.40)\times10^{-3}_{[\bar{K}^{*0}\to\pi^+K^-]}$} & \XCResultCol{$(9.56\pm3.37)\times10^{-4}_{[\bar{K}^{*0}\to\pi^+K^-]}$} \\
		\hline
		\XCModeCol{$\chi_{c0}\to \pi^0K^{0}\bar{K}^{*0}$} & \XCResultCol{$(5.65\pm1.99)\times10^{-4}_{[K^{*0}\to\pi^0K^{0}]}$} & \XCResultCol{$(4.78\pm1.69)\times10^{-4}_{[K^{*0}\to\pi^0K^{0}]}$} \\
		\hline
		\XCModeCol{$\chi_{c0}\to \pi^0\bar{K}^{0}K^{*0}$} & \XCResultCol{$(5.65\pm1.99)\times10^{-4}_{[\bar{K}^{*0}\to\pi^0\bar{K}^{0}]}$} & \XCResultCol{$(4.78\pm1.69)\times10^{-4}_{[\bar{K}^{*0}\to\pi^0\bar{K}^{0}]}$} \\
		\hline
		\multirow{2}{*}{\XCModeCol{$\chi_{c0}\to \pi^+\pi^-\omega$}} & \XCResultCol{$(1.50\pm0.28)\times10^{-5}_{[\omega\to\pi^+\pi^-]}$} & \XCResultCol{$(1.26\pm0.24)\times10^{-5}_{[\omega\to\pi^+\pi^-]}$} \\
		& \XCResultCol{$(1.37\pm0.39)\times10^{-8}_{[\phi\to\pi^+\pi^-]}$} & \XCResultCol{$(1.16\pm0.33)\times10^{-8}_{[\phi\to\pi^+\pi^-]}$} \\
		\hline
		\multirow{2}{*}{\XCModeCol{$\chi_{c0}\to \pi^0\pi^0\omega$}} & \XCResultCol{$<2.36\times10^{-7}_{[\omega\to\pi^0\pi^0]}$} & \XCResultCol{$<2.00\times10^{-7}_{[\omega\to\pi^0\pi^0]}$} \\
		& \XCResultCol{$(6.92\pm1.94)\times10^{-9}_{[\phi\to\pi^0\pi^0]}$} & \XCResultCol{$(5.84\pm1.65)\times10^{-9}_{[\phi\to\pi^0\pi^0]}$} \\
		\hline
		\multirow{2}{*}{\XCModeCol{$\chi_{c0}\to \pi^+\pi^-\phi$}} & \XCResultCol{$(2.19\pm0.36)\times10^{-6}_{[\omega\to\pi^+\pi^-]}$} & \XCResultCol{$(1.85\pm0.31)\times10^{-6}_{[\omega\to\pi^+\pi^-]}$} \\
		& \XCResultCol{$(8.12\pm1.89)\times10^{-8}_{[\phi\to\pi^+\pi^-]}$} & \XCResultCol{$(6.84\pm1.60)\times10^{-8}_{[\phi\to\pi^+\pi^-]}$} \\
		\hline
		\multirow{2}{*}{\XCModeCol{$\chi_{c0}\to \pi^0\pi^0\phi$}} & \XCResultCol{$<3.40\times10^{-8}_{[\omega\to\pi^0\pi^0]}$} & \XCResultCol{$<2.86\times10^{-8}_{[\omega\to\pi^0\pi^0]}$} \\
		& \XCResultCol{$(4.10\pm0.95)\times10^{-8}_{[\phi\to\pi^0\pi^0]}$} & \XCResultCol{$(3.45\pm0.79)\times10^{-8}_{[\phi\to\pi^0\pi^0]}$} \\
		\hline
		\XCModeCol{$\chi_{c0}\to K^-K^+\omega$} & \XCResultCol{$(7.12\pm0.73)\times10^{-5}_{[\phi\to K^-K^+]}$} & \XCResultCol{$(6.03\pm0.61)\times10^{-5}_{[\phi\to K^-K^+]}$} \\
		\hline
		\XCModeCol{$\chi_{c0}\to K^{0}\bar{K}^{0}\omega$} & \XCResultCol{$(4.77\pm0.49)\times10^{-5}_{[\phi\to K^{0}\bar{K}^{0}]}$} & \XCResultCol{$(4.05\pm0.41)\times10^{-5}_{[\phi\to K^{0}\bar{K}^{0}]}$} \\
		\hline
		\XCModeCol{$\chi_{c0}\to K^-K^+\phi$} & \XCResultCol{$(4.25\pm0.20)\times10^{-4}_{[\phi\to K^-K^+]}$} & \XCResultCol{$(3.60\pm0.17)\times10^{-4}_{[\phi\to K^-K^+]}$} \\
		\hline
		\XCModeCol{$\chi_{c0}\to K^{0}\bar{K}^{0}\phi$} & \XCResultCol{$(2.85\pm0.14)\times10^{-4}_{[\phi\to K^{0}\bar{K}^{0}]}$} & \XCResultCol{$(2.42\pm0.12)\times10^{-4}_{[\phi\to K^{0}\bar{K}^{0}]}$} \\
		\hline
	\end{tabular}
	\endgroup
	\label{Tab:B2PVlv1}
\end{table}

\begin{sidewaystable}[p]
	\centering
	\renewcommand\arraystretch{0.96}
	\setlength{\tabcolsep}{1.1pt}
	\caption{Branching ratios for the $\chi_{c1}\to P_1P_2V$ decays arising from vector- and tensor-meson resonance contributions (Part I). NWA denotes results under the narrow-width approximation, whereas FWE denotes results with finite-width effects included. A $\#$ mark identifies a tensor subdecay for which the NWA is not applicable.}
	\label{Tab:Brchi12PPVpartI}
	\begin{adjustbox}{max width=0.99\textwidth,max totalheight=0.86\textheight,keepaspectratio}
		\begin{tabular}{l|l|ll|ll}
			\hline\hline
			Decay mode & Experimental data & $T$-resonance contribution (NWA) & $T$-resonance contribution (FWE) & $V$-resonance contribution (NWA) & $V$-resonance contribution (FWE) \\
			\hline
			$\chi_{c1}\to \pi^0K^+K^{*-}$ & $\cdots$ & $(7.05\pm1.94)\times10^{-5}_{[K_2^*(1430)^-\to\pi^0K^{*-}]}$ & $(6.74\pm1.85)\times10^{-5}_{[K_2^*(1430)^-\to\pi^0K^{*-}]}$ & $\cdots$   &  $\cdots$  \\
			&    &    & $(2.36\pm0.62)\times10^{-7}_{[a_2(1320)^0\to K^+K^{*-}]\OffShellMark}$ &    &    \\
			\hline
			$\chi_{c1}\to \pi^0K^-K^{*+}$	& $\cdots$   &    & $(2.35\pm0.62)\times10^{-7}_{[a_2(1320)^0\to K^-K^{*+}]\OffShellMark}$ & $(4.40\pm1.07)\times10^{-4}_{[K^{*-}\to\pi^0K^-]}$ & $(3.76\pm1.06)\times10^{-4}_{[K^{*-}\to\pi^0K^-]}$ \\
			\hline
			$\chi_{c1}\to\pi^0\bar{K}^{0}K^{*0}$ & $\cdots$ &  & $(1.95\pm0.54)\times10^{-7}_{[a_2(1320)^0\to \bar{K}^{0}K^{*0}]\OffShellMark}$ & $(4.39\pm1.06)\times10^{-4}_{[\bar{K}^{*0}\to\pi^0\bar{K}^{0}]}$ & $(3.71\pm0.90)\times10^{-4}_{[\bar{K}^{*0}\to\pi^0\bar{K}^{0}]}$ \\
			\hline
			$\chi_{c1}\to \pi^-K^+\bar{K}^{*0}$ & $(1.6\pm0.9)\times10^{-3}$ & $(1.36\pm0.37)\times10^{-4}_{[K_2^*(1430)^-\to \pi^-\bar{K}^{*0}]}$ & $(1.30\pm0.36)\times10^{-4}_{[K_2^*(1430)^-\to \pi^-\bar{K}^{*0}]}$ & $(8.78\pm2.13)\times10^{-4}_{[K^{*0}\to\pi^-K^+]}$ & $(7.42\pm1.80)\times10^{-4}_{[K^{*0}\to\pi^-K^+]}$ \\
			\hline
			$\chi_{c1}\to \pi^+K^-K^{*0}$ & $\cdots$ &      & $(4.24\pm1.13)\times10^{-7}_{[a_2(1320)^-\to K^-K^{*0}]\OffShellMark}$ & $(8.78\pm2.13)\times10^{-4}_{[\bar{K}^{*0}\to\pi^+K^-]}$ & $(7.42\pm1.80)\times10^{-4}_{[\bar{K}^{*0}\to\pi^+K^-]}$ \\
			\hline
			$\chi_{c1}\to K^-K^+\omega$ & $(7.3\pm0.8)\times10^{-4}$ & $(2.55\pm1.19)\times10^{-5}_{[K_2^*(1430)^-\to K^-\omega]}$ & $(3.23\pm1.55)\times10^{-5}_{[K_2^*(1430)^-\to K^-\omega]}$ & $(1.36\pm0.22)\times10^{-5}_{[\phi\to K^-K^+]}$ & $(1.15\pm0.18)\times10^{-5}_{[\phi\to K^-K^+]}$ \\
			\hline
			$\chi_{c1}\to\pi^+\pi^-\omega$ & $\cdots$ & $\cdots$ & $\cdots$ & $(8.79\pm1.73)\times10^{-6}_{[\omega\to\pi^+\pi^-]}$ & $(7.43\pm1.45)\times10^{-6}_{[\omega\to\pi^+\pi^-]}$ \\
			&    &    &    & $(2.65\pm0.88)\times10^{-9}_{[\phi\to\pi^+\pi^-]}$ & $(2.23\pm0.74)\times10^{-9}_{[\phi\to\pi^+\pi^-]}$ \\
			\hline
			$\chi_{c1}\to\pi^0\pi^0\omega$ & $\cdots$ & $\cdots$ & $\cdots$ & $<1.40\times10^{-7}_{[\omega\to\pi^0\pi^0]}$ & $<1.18\times10^{-7}_{[\omega\to\pi^0\pi^0]}$ \\
			&    &    &    & $(1.33\pm0.44)\times10^{-9}_{[\phi\to\pi^0\pi^0]}$ & $(1.12\pm0.37)\times10^{-9}_{[\phi\to\pi^0\pi^0]}$ \\
			\hline
			$\chi_{c1}\to\pi^+\pi^-\phi$ & $\cdots$ & $\cdots$ & $\cdots$ & $(4.18\pm0.92)\times10^{-7}_{[\omega\to\pi^+\pi^-]}$ & $(3.52\pm0.78)\times10^{-7}_{[\omega\to\pi^+\pi^-]}$ \\
			&    &    &    & $(4.09\pm0.99)\times10^{-8}_{[\phi\to\pi^+\pi^-]}$ & $(3.45\pm0.84)\times10^{-8}_{[\phi\to\pi^+\pi^-]}$ \\
			\hline
			$\chi_{c1}\to\pi^0\pi^0\phi$ & $\cdots$ & $\cdots$ & $\cdots$ & $<6.78\times10^{-9}_{[\omega\to\pi^0\pi^0]}$ & $<5.76\times10^{-9}_{[\omega\to\pi^0\pi^0]}$ \\
			&    &    &    & $(2.06\pm0.50)\times10^{-8}_{[\phi\to\pi^0\pi^0]}$ & $(1.74\pm0.42)\times10^{-8}_{[\phi\to\pi^0\pi^0]}$ \\
			\hline
			$\chi_{c1}\to K^-K^+\phi$ & $(4.1\pm1.5)\times10^{-4}$ &  &  $(3.42\pm2.74)\times10^{-7}_{[K_2^*(1430)^-\to K^-\phi]\OffShellMark}$ & $(2.14\pm0.13)\times10^{-4}_{[\phi\to K^-K^+]}$ & $(1.81\pm0.11)\times10^{-4}_{[\phi\to K^-K^+]}$ \\
			\hline
			$\chi_{c1}\to K^{0}\bar{K}^{0}\phi$ & $\cdots$ &  & $(3.89\pm3.17)\times10^{-7}_{[\bar{K}_2^{*}(1430)^0\to \bar{K}^{0}\phi]\OffShellMark}$ & $(1.43\pm0.09)\times10^{-4}_{[\phi\to K^{0}\bar{K}^{0}]}$ & $(1.21\pm0.07)\times10^{-4}_{[\phi\to K^{0}\bar{K}^{0}]}$ \\
			\hline
			$\chi_{c1}\to \pi^-\bar{K}^{0}K^{*+}$ & $\cdots$ & $\cdots$ & $\cdots$ & $(8.80\pm2.14)\times10^{-4}_{[K^{*-}\to \bar{K}^{0}\pi^-]}$ & $(7.52\pm2.12)\times10^{-4}_{[K^{*-}\to \bar{K}^{0}\pi^-]}$ \\
			\hline
			$\chi_{c1}\to \pi^+K^0K^{*-}$ & $\cdots$ & $(9.73\pm2.52)\times10^{-5}_{[\bar{K}_2^{*}(1430)^0\to \pi^+K^{*-}]}$ & $(9.47\pm2.51)\times10^{-5}_{[\bar{K}_2^{*}(1430)^0\to \pi^+K^{*-}]}$ & $\cdots$ & $\cdots$ \\
			&    &    & $(4.30\pm1.14)\times10^{-7}_{[a_2(1320)^-\to K^0K^{*-}]\OffShellMark}$ &    &    \\
			\hline
			$\chi_{c1}\to \pi^0K^{0}\bar{K}^{*0}$ & $\cdots$ & $(4.79\pm1.24)\times10^{-5}_{[\bar{K}_2^{*}(1430)^0\to \pi^0\bar{K}^{*0}]}$ & $(4.68\pm1.23)\times10^{-5}_{[\bar{K}_2^{*}(1430)^0\to \pi^0\bar{K}^{*0}]}$ & $(4.39\pm1.06)\times10^{-4}_{[K^{*0}\to\pi^0K^{0}]}$ & $(3.71\pm0.90)\times10^{-4}_{[K^{*0}\to\pi^0K^{0}]}$ \\
			&    &    & $(1.95\pm0.54)\times10^{-7}_{[a_2(1320)^0\to K^0\bar{K}^{*0}]\OffShellMark}$ &    &    \\
			\hline
			$\chi_{c1}\to K^{0}\bar{K}^{0}\omega$ & $\cdots$ & $(1.75\pm0.78)\times10^{-5}_{[\bar{K}_2^{*}(1430)^0\to \bar{K}^{0}\omega]}$ & $(2.33\pm1.09)\times10^{-5}_{[\bar{K}_2^{*}(1430)^0\to \bar{K}^{0}\omega]}$ & $(9.09\pm1.44)\times10^{-6}_{[\phi\to K^{0}\bar{K}^{0}]}$ & $(7.71\pm1.23)\times10^{-6}_{[\phi\to K^{0}\bar{K}^{0}]}$ \\
			\hline
		\end{tabular}
	\end{adjustbox}
\end{sidewaystable}

\begin{sidewaystable}[p]
	\centering
	\renewcommand\arraystretch{0.80}
	\setlength{\tabcolsep}{1.4pt}
	\caption{Branching ratios for the $\chi_{c1}\to P_1P_2V$ decays arising from vector- and tensor-meson resonance contributions (Part II). NWA denotes results under the narrow-width approximation, whereas FWE denotes results with finite-width effects included. A $\#$ mark identifies a tensor subdecay for which the NWA is not applicable.}
	\label{Tab:Brchi12PPVpartII}
	\begin{adjustbox}{max width=0.94\textwidth,max totalheight=0.78\textheight,keepaspectratio,center}
		\begin{tabular}{l|ll|ll}
			\hline\hline
			Decay mode & $T$-resonance contribution (NWA) & $T$-resonance contribution (FWE) & $V$-resonance contribution (NWA) & $V$-resonance contribution (FWE) \\
			\hline
			$\chi_{c1}\to \pi^-\pi^+\rho^0$ & $(2.48\pm0.17)\times10^{-4}_{[a_2(1320)^-\to\pi^-\rho^0]}$ & $(2.40\pm0.19)\times10^{-4}_{[a_2(1320)^-\to\pi^-\rho^0]}$ & $(7.51\pm2.46)\times10^{-4}_{[\rho^0\to\pi^+\pi^-]}$ & $(6.29\pm2.06)\times10^{-4}_{[\rho^0\to\pi^+\pi^-]}$ \\
			\hline
			{$\chi_{c1}\to \pi^0\pi^+\rho^-$} & $(2.52\pm0.18)\times10^{-4}_{[a_2(1320)^-\to\pi^0\rho^-]}$ & $(2.43\pm0.19)\times10^{-4}_{[a_2(1320)^-\to\pi^0\rho^-]}$ &  $\cdots$  & $\cdots$   \\
			& $(2.48\pm0.17)\times10^{-4}_{[a_2(1320)^0\to\pi^+\rho^-]}$ & $(2.41\pm0.19)\times10^{-4}_{[a_2(1320)^0\to\pi^+\rho^-]}$ &    &    \\	\hline
			{$\chi_{c1}\to \pi^0\pi^-\rho^+$}& $(2.49\pm0.18)\times10^{-4}_{[a_2(1320)^0\to\pi^-\rho^+]}$ & $(2.41\pm0.19)\times10^{-4}_{[a_2(1320)^0\to\pi^-\rho^+]}$ & $(1.52\pm0.50)\times10^{-3}_{[\rho^-\to\pi^-\pi^0]}$ & $(1.27\pm0.42)\times10^{-3}_{[\rho^-\to\pi^-\pi^0]}$ \\
			\hline
			
			$\chi_{c1}\to K^-K^+\rho^0$ & $(2.47\pm0.77)\times10^{-5}_{[K_2^*(1430)^-\to K^-\rho^0]}$ & $(3.02\pm0.97)\times10^{-5}_{[K_2^*(1430)^-\to K^-\rho^0]}$ & $\cdots$ & $\cdots$ \\
			\hline
			$\chi_{c1}\to \bar{K}^{0}K^+\rho^-$ & $(4.68\pm1.46)\times10^{-5}_{[K_2^*(1430)^-\to \bar{K}^{0}\rho^-]}$ & $(5.78\pm1.84)\times10^{-5}_{[K_2^*(1430)^-\to \bar{K}^{0}\rho^-]}$ & $\cdots$ & $\cdots$ \\
			\hline
			$\chi_{c1}\to K^0K^-\rho^+$ & $(3.57\pm1.07)\times10^{-5}_{[\bar{K}_2^{*}(1430)^0\to K^-\rho^+]}$ & $(4.57\pm1.46)\times10^{-5}_{[\bar{K}_2^{*}(1430)^0\to K^-\rho^+]}$ & $\cdots$ & $\cdots$ \\
			\hline
			$\chi_{c1}\to \bar{K}^{0}K^0\rho^0$ & $(1.69\pm0.50)\times10^{-5}_{[\bar{K}_2^{*}(1430)^0\to \bar{K}^{0}\rho^0]}$ & $(2.19\pm0.69)\times10^{-5}_{[\bar{K}_2^{*}(1430)^0\to \bar{K}^{0}\rho^0]}$ & $\cdots$ & $\cdots$ \\
			\hline
			$\chi_{c1}\to \eta K^-K^{*+}$ & $(1.24\pm0.40)\times10^{-5}_{[f'_2(1525)\to K^-K^{*+}]}$ & $(1.40\pm0.49)\times10^{-5}_{[f'_2(1525)\to K^-K^{*+}]}$ & $\cdots$ & $\cdots$ \\
			&    & $(3.79\pm1.21)\times10^{-7}_{[f_2(1270)\to K^-K^{*+}]\OffShellMark}$ &    &    \\
			\hline
			& $(1.23\pm0.39)\times10^{-5}_{[f'_2(1525)\to K^+K^{*-}]}$ & $(1.40\pm0.49)\times10^{-5}_{[f'_2(1525)\to K^+K^{*-}]}$ & $\cdots$ & $\cdots$ \\
			$\chi_{c1}\to \eta K^+K^{*-}$&    & $(3.79\pm1.21)\times10^{-7}_{[f_2(1270)\to K^+K^{*-}]\OffShellMark}$ &    &    \\
			&    & $(3.21\pm1.78)\times10^{-6}_{[K_2^*(1430)^-\to \eta K^{*-}]\OffShellMark}$ &    &    \\
			\hline
			$\chi_{c1}\to \eta \bar{K}^{0}K^{*0}$ & $(1.07\pm0.35)\times10^{-5}_{[f'_2(1525)\to \bar{K}^{0}K^{*0}]}$ & $(1.26\pm0.46)\times10^{-5}_{[f'_2(1525)\to \bar{K}^{0}K^{*0}]}$ & $\cdots$ & $\cdots$ \\
			&    & $(3.37\pm1.09)\times10^{-7}_{[f_2(1270)\to \bar{K}^{0}K^{*0}]\OffShellMark}$ &    &    \\
			\hline
			$\chi_{c1}\to \eta K^{0}\bar{K}^{*0}$ & $(1.07\pm0.34)\times10^{-5}_{[f'_2(1525)\to K^{0}\bar{K}^{*0}]}$ & $(1.27\pm0.45)\times10^{-5}_{[f'_2(1525)\to K^{0}\bar{K}^{*0}]}$ & $\cdots$ & $\cdots$ \\
			&    & $(3.37\pm1.09)\times10^{-7}_{[f_2(1270)\to K^{0}\bar{K}^{*0}]\OffShellMark}$ &    &    \\
			&    & $(2.85\pm1.52)\times10^{-6}_{[\bar{K}_2^{*}(1430)^0\to\eta\bar{K}^{*0}]\OffShellMark}$ &    &    \\
			\hline
			$\chi_{c1}\to\eta\pi^0\rho^0$ & $\cdots$ & $\cdots$ & $<5.97\times10^{-6}_{[\rho^0\to\pi^0\eta]}$ &
			$<7.31\times10^{-6}_{[\rho^0\to\pi^0\eta]}$ \\
			\hline
			$\chi_{c1}\to\eta\pi^-\rho^+$ & $\cdots$ & $\cdots$ & $<1.19\times10^{-5}_{[\rho^-\to\pi^-\eta]}$ & $<1.52\times10^{-5}_{[\rho^-\to\pi^-\eta]}$ \\
			\hline
			$\chi_{c1}\to \eta^{\prime}K^-K^{*+}$ & $(3.21\pm0.99)\times10^{-5}_{[f'_2(1525)\to K^-K^{*+}]}$ & $(3.64\pm1.22)\times10^{-5}_{[f'_2(1525)\to K^-K^{*+}]}$ & $\cdots$ & $\cdots$ \\
			&    & $(5.49\pm1.98)\times10^{-7}_{[f_2(1270)\to K^-K^{*+}]\OffShellMark}$ &    &    \\
			\hline
			$\chi_{c1}\to \eta^{\prime}K^+K^{*-}$ & $(3.21\pm0.99)\times10^{-5}_{[f'_2(1525)\to K^+K^{*-}]}$ & $(3.62\pm1.22)\times10^{-5}_{[f'_2(1525)\to K^+K^{*-}]}$ & $\cdots$ & $\cdots$ \\
			&    & $(5.50\pm2.00)\times10^{-7}_{[f_2(1270)\to K^+K^{*-}]\OffShellMark}$ &    &    \\
			\hline
			$\chi_{c1}\to \eta^{\prime}\bar{K}^{0}K^{*0}$ & $(2.79\pm0.87)\times10^{-5}_{[f'_2(1525)\to \bar{K}^{0}K^{*0}]}$ & $(3.23\pm1.12)\times10^{-5}_{[f'_2(1525)\to \bar{K}^{0}K^{*0}]}$ & $\cdots$ & $\cdots$ \\
			&    & $(4.88\pm1.79)\times10^{-7}_{[f_2(1270)\to \bar{K}^{0}K^{*0}]\OffShellMark}$ &    &    \\
			\hline
			$\chi_{c1}\to \eta^{\prime}K^{0}\bar{K}^{*0}$ & $(2.79\pm0.87)\times10^{-5}_{[f'_2(1525)\to K^{0}\bar{K}^{*0}]}$ & $(3.26\pm1.13)\times10^{-5}_{[f'_2(1525)\to K^{0}\bar{K}^{*0}]}$ & $\cdots$ & $\cdots$ \\
			&    & $(4.88\pm1.79)\times10^{-7}_{[f_2(1270)\to K^{0}\bar{K}^{*0}]\OffShellMark}$ &    &    \\
			\hline
		\end{tabular}
	\end{adjustbox}
\end{sidewaystable}

\begin{sidewaystable}[p]
	\centering
	\renewcommand\arraystretch{0.96}
	\setlength{\tabcolsep}{1.1pt}
	\caption{Branching ratios for the $\chi_{c2}\to P_1P_2V$ decays arising from vector- and tensor-meson resonance contributions (Part I). NWA denotes results under the narrow-width approximation, whereas FWE denotes results with finite-width effects included. A $\#$ mark identifies a tensor subdecay for which the NWA is not applicable.}
	\label{Tab:Brchi22PPVpartI}
	\begin{adjustbox}{max width=0.99\textwidth,max totalheight=0.86\textheight,keepaspectratio}
		\begin{tabular}{l|l|ll|ll}
			\hline\hline
			Decay mode & Experimental data & $T$-resonance contribution (NWA) & $T$-resonance contribution (FWE) & $V$-resonance contribution (NWA) & $V$-resonance contribution (FWE) \\
			\hline
			$\chi_{c2}\to \pi^0K^+K^{*-}$ & $\cdots$ & $(6.56\pm1.14)\times10^{-5}_{[K_2^*(1430)^-\to\pi^0K^{*-}]}$ & $(6.27\pm1.10)\times10^{-5}_{[K_2^*(1430)^-\to\pi^0K^{*-}]}$ &  $\cdots$  &  $\cdots$  \\
			&    &    & $(4.61\pm2.17)\times10^{-7}_{[a_2(1320)^0\to K^+K^{*-}]\OffShellMark}$ &    &    \\
			\hline
			$\chi_{c2}\to \pi^0K^-K^{*+}$ & $\cdots$   &    & $(4.59\pm2.16)\times10^{-7}_{[a_2(1320)^0\to K^-K^{*+}]\OffShellMark}$ & $(6.79\pm2.45)\times10^{-4}_{[K^{*-}\to\pi^0K^-]}$ & $(5.82\pm2.32)\times10^{-4}_{[K^{*-}\to\pi^0K^-]}$ \\
			\hline
			$\chi_{c2}\to\pi^0\bar{K}^{0}K^{*0}$ & $\cdots$ &  & $(3.82\pm1.83)\times10^{-7}_{[a_2(1320)^0\to \bar{K}^{0}K^{*0}]\OffShellMark}$ & $(6.77\pm2.45)\times10^{-4}_{[\bar{K}^{*0}\to\pi^0\bar{K}^{0}]}$ & $(5.72\pm2.07)\times10^{-4}_{[\bar{K}^{*0}\to\pi^0\bar{K}^{0}]}$ \\
			\hline
			$\chi_{c2}\to \pi^-K^+\bar{K}^{*0}$ & $(2.1\pm1.1)\times10^{-3}$ & $(1.26\pm0.22)\times10^{-4}_{[K_2^*(1430)^-\to \pi^-\bar{K}^{*0}]}$ & $(1.20\pm0.21)\times10^{-4}_{[K_2^*(1430)^-\to \pi^-\bar{K}^{*0}]}$ & $(1.35\pm0.49)\times10^{-3}_{[K^{*0}\to\pi^-K^+]}$ & $(1.14\pm0.41)\times10^{-3}_{[K^{*0}\to\pi^-K^+]}$ \\
			\hline
			$\chi_{c2}\to \pi^+K^-K^{*0}$ & $\cdots$ &       & $(5.78\pm3.04)\times10^{-7}_{[a_2(1320)^-\to K^-K^{*0}]\OffShellMark}$ & $(1.35\pm0.49)\times10^{-3}_{[\bar{K}^{*0}\to\pi^+K^-]}$ & $(1.14\pm0.41)\times10^{-3}_{[\bar{K}^{*0}\to\pi^+K^-]}$ \\
			\hline
			$\chi_{c2}\to K^-K^+\omega$ & $(6.6\pm0.8)\times10^{-4}$ &
			$(2.32\pm0.89)\times10^{-5}_{[K_2^*(1430)^-\to K^-\omega]}$ & $(2.94\pm1.15)\times10^{-5}_{[K_2^*(1430)^-\to K^-\omega]}$ & $(4.88\pm1.45)\times10^{-6}_{[\phi\to K^-K^+]}$ & $(4.12\pm1.23)\times10^{-6}_{[\phi\to K^-K^+]}$ \\
			\hline
			$\chi_{c2}\to\pi^+\pi^-\omega$ & $\cdots$ & $\cdots$ & $\cdots$ & $(1.33\pm0.25)\times10^{-5}_{[\omega\to\pi^+\pi^-]}$ & $(1.12\pm0.21)\times10^{-5}_{[\omega\to\pi^+\pi^-]}$ \\
			&    &    &    & $(9.77\pm4.47)\times10^{-10}_{[\phi\to\pi^+\pi^-]}$ & $(8.21\pm3.74)\times10^{-10}_{[\phi\to\pi^+\pi^-]}$ \\
			\hline
			$\chi_{c2}\to\pi^0\pi^0\omega$ & $\cdots$ & $\cdots$ & $\cdots$ & $<2.10\times10^{-7}_{[\omega\to\pi^0\pi^0]}$ & $<1.77\times10^{-7}_{[\omega\to\pi^0\pi^0]}$ \\
			&    &    &    & $(4.88\pm2.23)\times10^{-10}_{[\phi\to\pi^0\pi^0]}$ & $(4.14\pm1.88)\times10^{-10}_{[\phi\to\pi^0\pi^0]}$ \\
			\hline
			$\chi_{c2}\to\pi^+\pi^-\phi$ & $\cdots$ & $\cdots$ & $\cdots$ & $(1.52\pm0.54)\times10^{-7}_{[\omega\to\pi^+\pi^-]}$ & $(1.28\pm0.45)\times10^{-7}_{[\omega\to\pi^+\pi^-]}$ \\
			&    &    &    & $(1.18\pm0.30)\times10^{-7}_{[\phi\to\pi^+\pi^-]}$ & $(9.98\pm2.52)\times10^{-8}_{[\phi\to\pi^+\pi^-]}$ \\
			\hline
			$\chi_{c2}\to\pi^0\pi^0\phi$ & $\cdots$ & $\cdots$ & $\cdots$ & $<2.72\times10^{-9}_{[\omega\to\pi^0\pi^0]}$ & $<2.28\times10^{-9}_{[\omega\to\pi^0\pi^0]}$ \\
			&    &    &    & $(5.96\pm1.48)\times10^{-8}_{[\phi\to\pi^0\pi^0]}$ & $(5.02\pm1.26)\times10^{-8}_{[\phi\to\pi^0\pi^0]}$ \\
			\hline
			$\chi_{c2}\to K^-K^+\phi$ & $(1.41\pm0.29)\times10^{-3}$ &  &  $(3.00\pm2.27)\times10^{-7}_{[K_2^*(1430)^-\to K^-\phi]\OffShellMark}$  & $(6.16\pm0.42)\times10^{-4}_{[\phi\to K^-K^+]}$ & $(5.22\pm0.36)\times10^{-4}_{[\phi\to K^-K^+]}$ \\
			\hline
			$\chi_{c2}\to K^{0}\bar{K}^{0}\phi$ & $\cdots$ &  & $(4.09\pm3.30)\times10^{-7}_{[\bar{K}_2^{*}(1430)^0\to \bar{K}^{0}\phi]\OffShellMark}$ & $(4.14\pm0.28)\times10^{-4}_{[\phi\to K^{0}\bar{K}^{0}]}$ & $(3.51\pm0.24)\times10^{-4}_{[\phi\to K^{0}\bar{K}^{0}]}$ \\
			\hline
			$\chi_{c2}\to \pi^-\bar{K}^{0}K^{*+}$ & $\cdots$ & $\cdots$ & $\cdots$ & $(1.36\pm0.49)\times10^{-3}_{[K^{*-}\to \bar{K}^{0}\pi^-]}$ & $(1.16\pm0.46)\times10^{-3}_{[K^{*-}\to \bar{K}^{0}\pi^-]}$ \\
			\hline
			$\chi_{c2}\to \pi^+K^0K^{*-}$ & $\cdots$ & $(1.05\pm0.24)\times10^{-4}_{[\bar{K}_2^{*}(1430)^0\to \pi^+K^{*-}]}$ & $(1.03\pm0.24)\times10^{-4}_{[\bar{K}_2^{*}(1430)^0\to \pi^+K^{*-}]}$ & $\cdots$ & $\cdots$ \\
			&    &    & $(5.87\pm3.06)\times10^{-7}_{[a_2(1320)^-\to K^0K^{*-}]\OffShellMark}$ &    &    \\
			\hline
			$\chi_{c2}\to \pi^0K^{0}\bar{K}^{*0}$ & $\cdots$ & $(5.18\pm1.17)\times10^{-5}_{[\bar{K}_2^{*}(1430)^0\to \pi^0\bar{K}^{*0}]}$ & $(5.07\pm1.17)\times10^{-5}_{[\bar{K}_2^{*}(1430)^0\to \pi^0\bar{K}^{*0}]}$ & $(6.77\pm2.45)\times10^{-4}_{[K^{*0}\to\pi^0K^{0}]}$ & $(5.72\pm2.07)\times10^{-4}_{[K^{*0}\to\pi^0K^{0}]}$ \\
			&    &    & $(3.81\pm1.84)\times10^{-7}_{[a_2(1320)^0\to K^0\bar{K}^{*0}]\OffShellMark}$ &    &    \\
			\hline
			$\chi_{c2}\to K^{0}\bar{K}^{0}\omega$ & $\cdots$ & $(1.89\pm0.80)\times10^{-5}_{[\bar{K}_2^{*}(1430)^0\to \bar{K}^{0}\omega]}$ & $(2.52\pm1.11)\times10^{-5}_{[\bar{K}_2^{*}(1430)^0\to \bar{K}^{0}\omega]}$ & $(3.27\pm0.98)\times10^{-6}_{[\phi\to K^{0}\bar{K}^{0}]}$ & $(2.77\pm0.82)\times10^{-6}_{[\phi\to K^{0}\bar{K}^{0}]}$ \\
			\hline
		\end{tabular}
	\end{adjustbox}
\end{sidewaystable}

\begin{sidewaystable}[p]
	\centering
	\renewcommand\arraystretch{0.80}
	\setlength{\tabcolsep}{1.4pt}
	\caption{Branching ratios for the $\chi_{c2}\to P_1P_2V$ decays arising from vector- and tensor-meson resonance contributions (Part II). NWA denotes results under the narrow-width approximation, whereas FWE denotes results with finite-width effects included. A $\#$ mark identifies a tensor subdecay for which the NWA is not applicable.}
	\label{Tab:Brchi22PPVpartII}
	\begin{adjustbox}{max width=0.94\textwidth,max totalheight=0.78\textheight,keepaspectratio,center}
		\begin{tabular}{l|ll|ll}
			\hline\hline
			Decay mode & $T$-resonance contribution (NWA) & $T$-resonance contribution (FWE) & $V$-resonance contribution (NWA) & $V$-resonance contribution (FWE) \\
			\hline
			$\chi_{c2}\to \pi^-\pi^+\rho^0$ & $(3.19\pm1.17)\times10^{-4}_{[a_2(1320)^-\to\pi^-\rho^0]}$ & $(3.09\pm1.16)\times10^{-4}_{[a_2(1320)^-\to\pi^-\rho^0]}$ & $(9.10\pm3.99)\times10^{-4}_{[\rho^0\to\pi^+\pi^-]}$ & $(7.61\pm3.33)\times10^{-4}_{[\rho^0\to\pi^+\pi^-]}$ \\
			\hline
			$\chi_{c2}\to \pi^0\pi^+\rho^-$
			& $(3.23\pm1.19)\times10^{-4}_{[a_2(1320)^-\to\pi^0\rho^-]}$ & $(3.13\pm1.17)\times10^{-4}_{[a_2(1320)^-\to\pi^0\rho^-]}$ & $\cdots$   & $\cdots$   \\
			& $(4.63\pm1.40)\times10^{-4}_{[a_2(1320)^0\to\pi^+\rho^-]}$ & $(4.49\pm1.40)\times10^{-4}_{[a_2(1320)^0\to\pi^+\rho^-]}$ &    &    \\ 	\hline
			$\chi_{c2}\to \pi^0\pi^-\rho^+$ & $(4.63\pm1.41)\times10^{-4}_{[a_2(1320)^0\to\pi^-\rho^+]}$ & $(4.50\pm1.40)\times10^{-4}_{[a_2(1320)^0\to\pi^-\rho^+]}$ & $(1.84\pm0.80)\times10^{-3}_{[\rho^-\to\pi^-\pi^0]}$ & $(1.54\pm0.68)\times10^{-3}_{[\rho^-\to\pi^-\pi^0]}$ \\
			\hline
			$\chi_{c2}\to K^-K^+\rho^0$ & $(2.29\pm0.49)\times10^{-5}_{[K_2^*(1430)^-\to K^-\rho^0]}$ & $(2.79\pm0.62)\times10^{-5}_{[K_2^*(1430)^-\to K^-\rho^0]}$ & $\cdots$ & $\cdots$ \\
			\hline
			$\chi_{c2}\to \bar{K}^{0}K^+\rho^-$ & $(4.32\pm0.92)\times10^{-5}_{[K_2^*(1430)^-\to \bar{K}^{0}\rho^-]}$ & $(5.37\pm1.19)\times10^{-5}_{[K_2^*(1430)^-\to \bar{K}^{0}\rho^-]}$ & $\cdots$ & $\cdots$ \\
			\hline
			$\chi_{c2}\to K^0K^-\rho^+$ & $(3.86\pm1.02)\times10^{-5}_{[\bar{K}_2^{*}(1430)^0\to K^-\rho^+]}$ & $(4.96\pm1.44)\times10^{-5}_{[\bar{K}_2^{*}(1430)^0\to K^-\rho^+]}$ & $\cdots$ & $\cdots$ \\
			\hline
			$\chi_{c2}\to \bar{K}^{0}K^0\rho^0$ & $(1.82\pm0.48)\times10^{-5}_{[\bar{K}_2^{*}(1430)^0\to \bar{K}^{0}\rho^0]}$ & $(2.35\pm0.68)\times10^{-5}_{[\bar{K}_2^{*}(1430)^0\to \bar{K}^{0}\rho^0]}$ & $\cdots$ & $\cdots$ \\
			\hline
			$\chi_{c2}\to \eta K^-K^{*+}$ & $(2.09\pm1.16)\times10^{-5}_{[f'_2(1525)\to K^-K^{*+}]}$ & $(2.39\pm1.38)\times10^{-5}_{[f'_2(1525)\to K^-K^{*+}]}$ & $\cdots$ & $\cdots$ \\
			&    & $(2.50\pm1.75)\times10^{-7}_{[f_2(1270)\to K^-K^{*+}]\OffShellMark}$ &    &    \\
			\hline
			& $(2.08\pm1.16)\times10^{-5}_{[f'_2(1525)\to K^+K^{*-}]}$ & $(2.39\pm1.37)\times10^{-5}_{[f'_2(1525)\to K^+K^{*-}]}$ & $\cdots$ & $\cdots$ \\
			$\chi_{c2}\to \eta K^+K^{*-}$&    & $(2.50\pm1.75)\times10^{-7}_{[f_2(1270)\to K^+K^{*-}]\OffShellMark}$ &    &    \\
			&    & $(2.89\pm1.38)\times10^{-6}_{[K_2^*(1430)^-\to \eta K^{*-}]\OffShellMark}$ &    &    \\
			\hline
			$\chi_{c2}\to \eta \bar{K}^{0}K^{*0}$ & $(1.81\pm1.02)\times10^{-5}_{[f'_2(1525)\to \bar{K}^{0}K^{*0}]}$ & $(2.14\pm1.24)\times10^{-5}_{[f'_2(1525)\to \bar{K}^{0}K^{*0}]}$ & $\cdots$ & $\cdots$ \\
			&    & $(2.22\pm1.55)\times10^{-7}_{[f_2(1270)\to \bar{K}^{0}K^{*0}]\OffShellMark}$ &    &    \\
			\hline
			$\chi_{c2}\to \eta K^{0}\bar{K}^{*0}$ & $(1.82\pm1.02)\times10^{-5}_{[f'_2(1525)\to K^{0}\bar{K}^{*0}]}$ & $(2.12\pm1.23)\times10^{-5}_{[f'_2(1525)\to K^{0}\bar{K}^{*0}]}$ & $\cdots$ & $\cdots$ \\
			&    & $(2.20\pm1.54)\times10^{-7}_{[f_2(1270)\to K^{0}\bar{K}^{*0}]\OffShellMark}$ &    &    \\
			&    & $(3.04\pm1.53)\times10^{-6}_{[\bar{K}_2^{*}(1430)^0\to\eta\bar{K}^{*0}]\OffShellMark}$ &    &    \\
			\hline
			$\chi_{c2}\to\eta\pi^0\rho^0$ & $\cdots$ & $\cdots$ & $<7.90\times10^{-6}_{[\rho^0\to\pi^0\eta]}$ &
			$<9.67\times10^{-6}_{[\rho^0\to\pi^0\eta]}$ \\
			\hline
			$\chi_{c2}\to\eta\pi^-\rho^+$ & $\cdots$ & $\cdots$ & $<1.58\times10^{-5}_{[\rho^-\to\pi^-\eta]}$ & $<1.98\times10^{-5}_{[\rho^-\to\pi^-\eta]}$ \\
			\hline
			$\chi_{c2}\to \eta^{\prime}K^-K^{*+}$ & $(2.26\pm2.04)\times10^{-5}_{[f'_2(1525)\to K^-K^{*+}]}$ & $(2.62\pm2.37)\times10^{-5}_{[f'_2(1525)\to K^-K^{*+}]}$ & $\cdots$ & $\cdots$ \\
			&    & $<8.72\times10^{-7}_{[f_2(1270)\to K^-K^{*+}]\OffShellMark}$ &    &    \\
			\hline
			$\chi_{c2}\to \eta^{\prime}K^+K^{*-}$ & $(2.25\pm2.04)\times10^{-5}_{[f'_2(1525)\to K^+K^{*-}]}$ & $(2.58\pm2.34)\times10^{-5}_{[f'_2(1525)\to K^+K^{*-}]}$ & $\cdots$ & $\cdots$ \\
			&    & $<8.74\times10^{-7}_{[f_2(1270)\to K^+K^{*-}]\OffShellMark}$ &    &    \\
			\hline
			$\chi_{c2}\to \eta^{\prime}\bar{K}^{0}K^{*0}$ & $(1.97\pm1.78)\times10^{-5}_{[f'_2(1525)\to \bar{K}^{0}K^{*0}]}$ & $(2.29\pm2.09)\times10^{-5}_{[f'_2(1525)\to \bar{K}^{0}K^{*0}]}$ & $\cdots$ & $\cdots$ \\
			&    & $<7.82\times10^{-7}_{[f_2(1270)\to \bar{K}^{0}K^{*0}]\OffShellMark}$ &    &    \\
			\hline
			$\chi_{c2}\to \eta^{\prime}K^{0}\bar{K}^{*0}$ & $(1.97\pm1.78)\times10^{-5}_{[f'_2(1525)\to K^{0}\bar{K}^{*0}]}$ & $(2.35\pm2.13)\times10^{-5}_{[f'_2(1525)\to K^{0}\bar{K}^{*0}]}$ & $\cdots$ & $\cdots$ \\
			&    & $<7.80\times10^{-7}_{[f_2(1270)\to K^{0}\bar{K}^{*0}]\OffShellMark}$ &    &    \\
			\hline
		\end{tabular}
	\end{adjustbox}
\end{sidewaystable}

\subsection{Dominant resonance patterns and data comparison}

Each resonance mechanism is evaluated separately, so the numbers quoted below characterize the magnitude of an individual intermediate-resonance band rather than a full decomposition of the three-body amplitude. This is precisely the quantity needed to identify which intermediate resonances are large enough to be resolved, and provide benchmark magnitudes and channel priorities for future amplitude analyses. Unless otherwise stated, the FWE results are quoted below.

The $\pi\pi\rho$ final states contain some of the largest resonance contributions for the three $\chi_{cJ}$ states. Since the $TP$ decay is absent for $\chi_{c0}$, only vector-meson resonances contribute to the $\chi_{c0}$ channels. The dominant bands are
\begin{align}
	B\bigl(\chi_{c0}\to\pi^{0}\pi^{-}\rho^{+}\bigr)
	&= (1.92\pm1.32)\times10^{-3}_{\quad[\rho^{-}\to\pi^{-}\pi^{0}]},\nonumber\\
	B\bigl(\chi_{c1}\to\pi^{0}\pi^{-}\rho^{+}\bigr)
	&= (1.27\pm0.42)\times10^{-3}_{\quad[\rho^{-}\to\pi^{-}\pi^{0}]},\nonumber\\
	B\bigl(\chi_{c1}\to\pi^{0}\pi^{-}\rho^{+}\bigr)
	&= (2.41\pm0.19)\times10^{-4}_{\quad[a_{2}(1320)^{0}\to\pi^{-}\rho^{+}]},\nonumber\\
	B\bigl(\chi_{c2}\to\pi^{0}\pi^{-}\rho^{+}\bigr)
	&= (1.54\pm0.68)\times10^{-3}_{\quad[\rho^{-}\to\pi^{-}\pi^{0}]},\nonumber\\
	B\bigl(\chi_{c2}\to\pi^{0}\pi^{-}\rho^{+}\bigr)
	&= (4.50\pm1.40)\times10^{-4}_{\quad[a_{2}(1320)^{0}\to\pi^{-}\rho^{+}]}.\nonumber
\end{align}
The vector band dominates in every case, while the $a_{2}(1320)$ tensor band remains sizable enough to be relevant in an amplitude analysis. These modes are natural first targets for experiment, given their comparatively large predicted rates.

The $\pi K\bar K^{*}$ modes form another important class, and for $\chi_{c0,1,2}\to\pi^{-}K^{+}\bar K^{*0}$ the measured totals allow a direct comparison:
\begin{align}
	B^{\mathrm{exp}}\bigl(\chi_{c0}\to\pi^{-}K^{+}\bar K^{*0}\bigr)
	&= (7.0\pm1.5)\times10^{-3},\nonumber\\
	B\bigl(\chi_{c0}\to\pi^{-}K^{+}\bar K^{*0}\bigr)
	&= (9.56\pm3.37)\times10^{-4}_{\quad[K^{*0}\to\pi^{-}K^{+}]},\nonumber\\
	B^{\mathrm{exp}}\bigl(\chi_{c1}\to\pi^{-}K^{+}\bar K^{*0}\bigr)
	&= (1.6\pm0.9)\times10^{-3},\nonumber\\
	B\bigl(\chi_{c1}\to\pi^{-}K^{+}\bar K^{*0}\bigr)
	&= (7.42\pm1.80)\times10^{-4}_{\quad[K^{*0}\to\pi^{-}K^{+}]},\nonumber\\
	B\bigl(\chi_{c1}\to\pi^{-}K^{+}\bar K^{*0}\bigr)
	&= (1.30\pm0.36)\times10^{-4}_{\quad[K^{*}_{2}(1430)^{-}\to\pi^{-}\bar K^{*0}]},\nonumber\\
	B^{\mathrm{exp}}\bigl(\chi_{c2}\to\pi^{-}K^{+}\bar K^{*0}\bigr)
	&= (2.1\pm1.1)\times10^{-3},\nonumber\\
	B\bigl(\chi_{c2}\to\pi^{-}K^{+}\bar K^{*0}\bigr)
	&= (1.14\pm0.41)\times10^{-3}_{\quad[K^{*0}\to\pi^{-}K^{+}]},\nonumber\\
	B\bigl(\chi_{c2}\to\pi^{-}K^{+}\bar K^{*0}\bigr)
	&= (1.20\pm0.21)\times10^{-4}_{\quad[K^{*}_{2}(1430)^{-}\to\pi^{-}\bar K^{*0}]}.\nonumber
\end{align}
For $\chi_{c0}$, the $K^{*0}$ band amounts to about $14\%$ of the measured total. For $\chi_{c1}$ and $\chi_{c2}$, the vector and tensor bands together are comparable to the measured totals---the resonance bands account for a large fraction of the rate, the remainder being attributable to nonresonant contributions, interference, and resonances not considered here. These comparisons carry physical content even without an interference treatment: they show that the $K^{*}$ and $K^{*}_{2}$ bands are individually large enough to be separated in an amplitude analysis, and that the vector band dominates the tensor band in the $\pi^{-}K^{+}\bar K^{*0}$ final states.

A similar pattern is found in the $\bar K K\phi$ channels,
\begin{align}
	B\bigl(\chi_{c0}\to K^{-}K^{+}\phi\bigr)
	&= (3.60\pm0.17)\times10^{-4}_{\quad[\phi\to K^{-}K^{+}]},\nonumber\\
	B^{\mathrm{exp}}\bigl(\chi_{c1}\to K^{-}K^{+}\phi\bigr)
	&= (4.1\pm1.5)\times10^{-4},\nonumber\\
	B\bigl(\chi_{c1}\to K^{-}K^{+}\phi\bigr)
	&= (1.81\pm0.11)\times10^{-4}_{\quad[\phi\to K^{-}K^{+}]},\nonumber\\
	B^{\mathrm{exp}}\bigl(\chi_{c2}\to K^{-}K^{+}\phi\bigr)
	&= (1.41\pm0.29)\times10^{-3},\nonumber\\
	B\bigl(\chi_{c2}\to K^{-}K^{+}\phi\bigr)
	&= (5.22\pm0.36)\times10^{-4}_{\quad[\phi\to K^{-}K^{+}]},\nonumber
\end{align}
where the $\phi$ band accounts for roughly $40\%$ of the measured rate in both $\chi_{c1}$ and $\chi_{c2}$, while the tensor band from $K^{*}_{2}(1430)^{-}\to\phi K^{-}$ is only of order $10^{-7}$ and hence negligible.
A robust pattern thus emerges: in the $\bar K K\phi$ final states a substantial part of the rate is carried by a single $\phi$ band, so an amplitude analysis of these modes can be performed with essentially one resonance, which strongly simplifies the fit and reduces the model dependence.

The $\bar K K\omega$ modes display the opposite hierarchy,
\begin{align}
	B^{\mathrm{exp}}\bigl(\chi_{c1}\to K^{-}K^{+}\omega\bigr)
	&= (7.3\pm0.8)\times10^{-4},\nonumber\\
	B\bigl(\chi_{c1}\to K^{-}K^{+}\omega\bigr)
	&= (3.23\pm1.55)\times10^{-5}_{\quad[K^{*}_{2}(1430)^{-}\to K^{-}\omega]},\nonumber\\
	B\bigl(\chi_{c1}\to K^{-}K^{+}\omega\bigr)
	&= (1.15\pm0.18)\times10^{-5}_{\quad[\phi\to K^{-}K^{+}]},\nonumber\\
	B^{\mathrm{exp}}\bigl(\chi_{c2}\to K^{-}K^{+}\omega\bigr)
	&= (6.6\pm0.8)\times10^{-4},\nonumber\\
	B\bigl(\chi_{c2}\to K^{-}K^{+}\omega\bigr)
	&= (2.94\pm1.15)\times10^{-5}_{\quad[K^{*}_{2}(1430)^{-}\to K^{-}\omega]},\nonumber\\
	B\bigl(\chi_{c2}\to K^{-}K^{+}\omega\bigr)
	&= (4.12\pm1.23)\times10^{-6}_{\quad[\phi\to K^{-}K^{+}]}.\nonumber
\end{align}
Here the $K^{*}_{2}(1430)$ tensor band exceeds the $\phi$ vector band by about a factor of three, so the $\bar K K\omega$ channels are the cleanest place to isolate the tensor resonance against a subleading vector background. We note, however, that the summed resonance bands remain well below the measured totals, indicating that a substantial nonresonant or additional-resonance component is present; a future amplitude analysis of these modes would directly test this expectation.

Overall, the resonance contributions organize the $P_{1}P_{2}V$ channels into three experimentally distinct classes: (i) $\pi\pi\rho$ and $\pi K\bar K^{*}$, where the $\rho$/$K^{*}$ vector band dominates but a comparable $a_{2}$/$K^{*}_{2}$ tensor band is present; (ii) $\bar K K\phi$, which is essentially a single $\phi$ band; and (iii) $\bar K K\omega$ and $\eta^{\prime}KK^{*}$, where the tensor band dominates and can be isolated. This classification, together with the magnitude ordering, provides a concrete priority list for future invariant-mass and amplitude analyses of the $\chi_{cJ}\to P_{1}P_{2}V$ decays.

\subsection{Off-shell contributions from below-threshold tensor subdecays}

The finite-width treatment is particularly important in the tensor sector because the $T\to PV$ rate has a strong $p^5$ dependence. The channels $K_2^*(1430)\to K^*\eta$, $K_2^*(1430)\to\phi K$, $a_2(1320)\to K^*K$, and $f_2(1270)\to K^*K$ are closed at the nominal tensor-meson mass. No contribution is obtained for these channels in the NWA. When the tensor-meson mass distribution is retained, the high-mass tail can reach the allowed region. Nonzero off-shell contributions can then be generated.

For most channels with sufficient phase space, the FWE values are moderately smaller than the corresponding NWA results in the present treatment. A different behavior can appear near thresholds because the phase space changes rapidly across the resonance distribution. Both the NWA and FWE results are therefore listed in tables.

For those kinematically closed tensor subdecays, the resulting contributions to the final $\chi_{cJ}\to P_1P_2V$ branching ratios are typically at the $10^{-7}$--$10^{-6}$ level. The $f'_2(1525)\to K^*K$ contributions in the $\eta'KK^*$ channels can reach the $10^{-5}$ level. Although these rates are not among the largest $P_1P_2V$ contributions, they are special in one respect: because these $T\to PV$ subdecays are kinematically closed at the nominal tensor-meson mass, the NWA is not applicable, and the entire contribution is generated by the high-mass tail of the finite-width distribution. These modes are therefore uniquely sensitive to the tensor-resonance line shape—its width and parametrization—rather than to the on-shell coupling probed by the open channels. A nonzero experimental signal in any of these modes would thus provide a stringent test of the finite-width treatment and of the resonance line shape adopted here.

\section{Conclusion}
\label{sec:conclusion}

We have studied the contributions of light vector- and tensor-meson resonances to the three-body decays $\chi_{cJ}(J=0,1,2)\to P_1P_2V$ through the intermediate processes $\chi_{c0,1,2}\to VV\to P_1P_2V$ and $\chi_{c1,2}\to TP\to P_1P_2V$. Within a common quasi-two-body framework, both the narrow-width approximation and the effects of a finite resonance width are treated consistently. The $P_1P_2V$ final states provide a direct window onto the $T\to PV$ tensor resonances—including their subthreshold contributions.

The predicted resonance contributions range from $\mathcal{O}(10^{-10})$ to $\mathcal{O}(10^{-3})$. The largest contributions are found in the $\pi\pi\rho$, $\pi K K^*$, and $K\bar K\phi$ channels, while the $K\bar K\omega$ and $\eta'KK^*$ modes provide additional sensitivity to tensor resonances. These channels fall into three classes: in the $\pi\pi\rho$ and $\pi KK^*$ modes the vector band dominates but a comparable tensor band is present; the $K^-K^+\phi$ modes are essentially a single $\phi$ band; and in the $K^-K^+\omega$ and $\eta'KK^*$ modes the tensor band dominates and can be isolated. It is worth noting that when the resonance mass lies close to or below the nominal threshold of a tensor subdecay, the NWA is not applicable, and the entire contribution arises from the high-mass tail of the finite-width distribution.

Each resonance contribution is computed separately, so the results quantify the individual intermediate-resonance bands and their relative importance across the $P_1P_2V$ final states; they are intended as priors and channel priorities for amplitude analyses rather than as a complete three-body amplitude. Interference among the different vector and tensor resonances, as well as possible nonresonant contributions, is not included; consequently, the comparison with a measured total branching ratio is a magnitude test of the resonance bands rather than a model-independent determination of the three-body rate.

These predictions can be tested experimentally by means of invariant-mass distributions, Dalitz-plot analyses, and full amplitude analyses, through which the $\rho$, $K^*$, $\phi$, $a_2$, $K_2^*$, and $f'_2$ bands and their relative phases can be examined. The $\pi\pi\rho$, $\pi K K^*$, and $K\bar K\phi$ channels are natural first choices, given their comparatively large predicted rates, whereas the $K\bar K\omega$ and $\eta'KK^*$ channels are useful for studying the tensor sector and finite-width effects.
\appendix
\section{SU(3) conventions and previously established ingredients}
\label{app:su3details}

This Appendix summarizes the light-quark mass decomposition and the meson multiplets used in the SU(3) flavor analysis. The diagonal light-quark mass matrix is decomposed as~\cite{Gronau:1995hm,Xu:2013dta,He:2014xha}
\begin{equation}
	\begin{pmatrix}
		m_u & 0 & 0\\
		0 & m_d & 0\\
		0 & 0 & m_s
	\end{pmatrix}
	=
	\frac{1}{3}(m_u+m_d+m_s)I
	+\frac{1}{2}(m_u-m_d)X
	+\frac{1}{6}(m_u+m_d-2m_s)W,
	\label{eq:quark_mass_decomposition}
\end{equation}
where $I$ is the identity matrix, while $X$ and $W$ are given by
\begin{equation}
	X=
	\begin{pmatrix}
		1&0&0\\
		0&-1&0\\
		0&0&0
	\end{pmatrix},
	\qquad
	W=
	\begin{pmatrix}
		1&0&0\\
		0&1&0\\
		0&0&-2
	\end{pmatrix}.
	\label{eq:XWmatrices}
\end{equation}
Only the $W$ term is retained to describe SU(3) flavor breaking. It represents the mass difference between the strange quark and the light $u$ and $d$ quarks.
The pseudoscalar-, vector-, and tensor-meson nonets in the SU(3) flavor basis $(u,d,s)$ are written as~\cite{Haber:1985cv,Mo:2024jjd}
\begin{align}
	P&=
	\begin{pmatrix}
		\frac{\pi^0}{\sqrt2}+\frac{\eta_8}{\sqrt6}+\frac{\eta_1}{\sqrt3}&\pi^+&K^+\\
		\pi^-&-\frac{\pi^0}{\sqrt2}+\frac{\eta_8}{\sqrt6}+\frac{\eta_1}{\sqrt3}&K^0\\
		K^-&\bar K^0&-\frac{2\eta_8}{\sqrt6}+\frac{\eta_1}{\sqrt3}
	\end{pmatrix},\label{eq:Pmatrix}\\[4pt]
	V&=
	\begin{pmatrix}
		\frac{\rho^0}{\sqrt2}+\frac{\omega_8}{\sqrt6}+\frac{\omega_1}{\sqrt3}&\rho^+&K^{*+}\\
		\rho^-&-\frac{\rho^0}{\sqrt2}+\frac{\omega_8}{\sqrt6}+\frac{\omega_1}{\sqrt3}&K^{*0}\\
		K^{*-}&\bar K^{*0}&-\frac{2\omega_8}{\sqrt6}+\frac{\omega_1}{\sqrt3}
	\end{pmatrix},\label{eq:Vmatrix}\\[4pt]
	T&=
	\begin{pmatrix}
		\frac{a_2^0}{\sqrt2}+\frac{f_2^8}{\sqrt6}+\frac{f_2^1}{\sqrt3}&a_2^+&K_2^{*+}\\
		a_2^-&-\frac{a_2^0}{\sqrt2}+\frac{f_2^8}{\sqrt6}+\frac{f_2^1}{\sqrt3}&K_2^{*0}\\
		K_2^{*-}&\bar K_2^{*0}&-\frac{2f_2^8}{\sqrt6}+\frac{f_2^1}{\sqrt3}
	\end{pmatrix}.
	\label{eq:Tmatrix}
\end{align}

Here, $\eta_1$ and $\eta_8$, $\omega_1$ and $\omega_8$, and $f_2^1$ and $f_2^8$ denote the unmixed SU(3) flavor-singlet and flavor-octet states in the pseudoscalar, vector, and tensor sectors, respectively. Following the mixing convention adopted by the PDG~\cite{PDG}, the physical states are defined by
\begin{alignat}{2}
	\eta &= \eta_8\cos\theta_P-\eta_1\sin\theta_P,\qquad &
	\eta' &= \eta_8\sin\theta_P+\eta_1\cos\theta_P,\label{eq:eta_mixing}\\[-2pt]
	\phi &= \omega_8\cos\theta_V-\omega_1\sin\theta_V,\qquad &
	\omega &= \omega_8\sin\theta_V+\omega_1\cos\theta_V,\label{eq:vector_mixing}\\[-2pt]
	f'_2 &= f_2^8\cos\theta_T-f_2^1\sin\theta_T,\qquad &
	f_2 &= f_2^8\sin\theta_T+f_2^1\cos\theta_T.\label{eq:tensor_mixing}
\end{alignat}
Here, $\theta_P$, $\theta_V$, and $\theta_T$ are the nonet mixing angles in the pseudoscalar, vector, and tensor sectors, respectively. In the numerical analysis, $\theta_P\in[-20^\circ,-10^\circ]$, $\theta_V=36.5^\circ$, and $\theta_T=(27\pm1)^\circ$ are used \cite{PDG}.

The generic SU(3) parameterizations used for the two-body subprocesses have been discussed previously~\cite{Lan:2025wkf,cc2PPP,T2PV}. No new derivation of these relations is intended in the present work.

The detailed vector-subdecay inputs used in the numerical analysis are listed here. The measured values are taken from the 2026 PDG~\cite{PDG}, while previously derived inputs are cited below. For the $V\to P_1P_2$ decays, the available experimental inputs are
\begin{equation}
	\begin{array}{@{}l@{\hspace{0.025\textwidth}}l@{}}
		\makebox[0.44\textwidth][l]{$\displaystyle \mathcal{B}(K^{*\pm}\to (K\pi)^\pm)=(99.896\pm0.010)\times10^{-2},$} &
		\makebox[0.44\textwidth][l]{$\displaystyle \mathcal{B}(K^{*0}\to (K\pi)^0)=(99.753\pm0.021)\times10^{-2},$}\\[3pt]
		\makebox[0.44\textwidth][l]{$\displaystyle \mathcal{B}(\rho^\pm\to\pi^\pm\pi^0)=(99.955\pm0.005)\times10^{-2},$} &
		\makebox[0.44\textwidth][l]{$\displaystyle \mathcal{B}(\rho^0\to\pi^+\pi^-)=(98.92\pm0.16)\times10^{-2},$}\\[3pt]
		\makebox[0.44\textwidth][l]{$\displaystyle \mathcal{B}(\phi\to K^+K^-)=(50.1\pm0.6)\times10^{-2},$} &
		\makebox[0.44\textwidth][l]{$\displaystyle \mathcal{B}(\omega\to\pi^+\pi^-)=(1.53\pm0.12)\times10^{-2},$}\\[3pt]
		\makebox[0.44\textwidth][l]{$\displaystyle \mathcal{B}(\omega\to\pi^0\pi^0)<2.2\times10^{-4},$} &
		\makebox[0.44\textwidth][l]{$\displaystyle \mathcal{B}(\rho^\pm\to\pi^\pm\eta)<6.0\times10^{-3},$}\\[3pt]
		\makebox[0.44\textwidth][l]{$\displaystyle \mathcal{B}(\phi\to\pi^+\pi^-)=(9.5\pm1.9)\times10^{-5},$} &
		\makebox[0.44\textwidth][l]{$\displaystyle \mathcal{B}(\phi\to K^0\bar K^0) = (33.6\pm0.4)\times10^{-2}.$}	
	\end{array}
\end{equation}
For unmeasured decay channels, the SU(3) predictions are used as inputs~\cite{cc2PPP}:
\begin{equation}
	\begin{array}{@{}l@{\hspace{0.025\textwidth}}l@{}}
		\makebox[0.44\textwidth][l]{$\displaystyle \mathcal{B}(\rho^0\to\pi^0\eta)  <  6.0\times10^{-3},$} &
		\makebox[0.44\textwidth][l]{$\displaystyle \mathcal{B}(\phi\to\pi^0\pi^0)  =  (4.79\pm0.95)\times10^{-5}.$}
	\end{array}
\end{equation}

\section{Updated $T\to PV$ amplitudes and inputs}
\label{app:t2pv}

The full channel amplitudes used in the present analysis are listed in Table~\ref{Tab:TtoPVamplitudes}. The symmetry-breaking terms are also displayed for completeness. The currently available $T\to PV$ data do not constrain the breaking coefficients with sufficient precision. Only the SU(3)-symmetric contributions are therefore retained in the numerical analysis.

Repeating the $T\to PV$ analysis of Ref.~\cite{T2PV} with updated input parameters, we obtain the tensor-subdecay inputs and predictions given in Table~\ref{Tab:TtoPVBR}.
\begin{table}[htbp]
\caption{SU(3) flavor amplitudes for the $T\to PV$ decays, including the symmetry-breaking terms.}
\centering
\begin{center}
\small
\renewcommand\arraystretch{1.06}\tabcolsep 1pt
\begin{tabular}{c|c}
\hline\hline
Decay mode & SU(3) flavor amplitude \\ \hline
$a_2^-\to \rho^0\pi^-$ & $-\sqrt{2}\,g_{10}-\sqrt{2}(g_{11}+g_{12})$ \\
$a_2^-\to \rho^-\pi^0$ & $\sqrt{2}\,g_{10}+\sqrt{2}(g_{11}+g_{12})$ \\
$a_2^0\to \rho^-\pi^+$ & $-\sqrt{2}\,g_{10}-\sqrt{2}(g_{11}+g_{12})$ \\
$a_2^0\to \rho^+\pi^-$ & $\sqrt{2}\,g_{10}+\sqrt{2}(g_{11}+g_{12})$ \\
\hline
$\bar K_2^{*0}\to K^{*-}\pi^+$ & $g_{10}+(g_{11}+g_{12})$ \\
$\bar K_2^{*0}\to \bar K^{*0}\pi^0$ & $-\frac{\sqrt{2}}{2}g_{10}-\frac{\sqrt{2}}{2}(g_{11}+g_{12})$ \\
$K_2^{*-}\to K^{*-}\pi^0$ & $\frac{\sqrt{2}}{2}g_{10}+\frac{\sqrt{2}}{2}(g_{11}+g_{12})$ \\
$K_2^{*-}\to \bar K^{*0}\pi^-$ & $g_{10}+(g_{11}+g_{12})$ \\
\hline
$K_2^{*-}\to \rho^0K^-$ & $-\frac{\sqrt{2}}{2}g_{10}-\frac{\sqrt{2}}{2}(g_{11}+g_{12})$ \\
$K_2^{*-}\to \rho^-\bar K^0$ & $-g_{10}-(g_{11}+g_{12})$ \\
$\bar K_2^{*0}\to \rho^+K^-$ & $-g_{10}-(g_{11}+g_{12})$ \\
$\bar K_2^{*0}\to \rho^0\bar K^0$ & $\frac{\sqrt{2}}{2}g_{10}+\frac{\sqrt{2}}{2}(g_{11}+g_{12})$ \\
\hline
$K_2^{*-}\to \omega K^-,\ \bar K_2^{*0}\to \omega\bar K^0$ & $-\frac{\sqrt{6}}{2}\sin\theta_V\,g_{10}+\sqrt{3}\cos\theta_V\,g_{20}+(-\frac{\sqrt{6}}{2}\sin\theta_V\,g_{11}-\sqrt{3}\cos\theta_V\,g_{12}+\frac{\sqrt{6}}{2}\sin\theta_V\,g_{12}+\sqrt{3}\cos\theta_V\,g_{21})$ \\
$K_2^{*-}\to \phi K^-,\ \bar K_2^{*0}\to \phi\bar K^0$ & $-\frac{\sqrt{6}}{2}\cos\theta_V\,g_{10}-\sqrt{3}\sin\theta_V\,g_{20}+(-\frac{\sqrt{6}}{2}\cos\theta_V\,g_{11}+\frac{\sqrt{6}}{2}\cos\theta_V\,g_{12}+\sqrt{3}\sin\theta_V\,g_{12}-\sqrt{3}\sin\theta_V\,g_{21})$ \\
$K_2^{*-}\to K^{*-}\eta,\ \bar K_2^{*0}\to \bar K^{*0}\eta$ & $\frac{\sqrt{6}}{2}\cos\theta_P\,g_{10}+\sqrt{3}\sin\theta_P\,g_{20}+(\frac{\sqrt{6}}{2}\cos\theta_P\,g_{11}-\frac{\sqrt{6}}{2}\cos\theta_P\,g_{12}-\sqrt{3}\sin\theta_P\,g_{12}+\sqrt{3}\sin\theta_P\,g_{21})$ \\
\hline
$a_2^0\to K^{*+}K^-$ & $\frac{\sqrt{2}}{2}g_{10}+(\frac{\sqrt{2}}{2}g_{11}-\sqrt{2}g_{12})$ \\
$a_2^0\to K^{*-}K^+$ & $-\frac{\sqrt{2}}{2}g_{10}+(-\frac{\sqrt{2}}{2}g_{11}+\sqrt{2}g_{12})$ \\
$a_2^0\to K^{*0}\bar K^0$ & $-\frac{\sqrt{2}}{2}g_{10}+(-\frac{\sqrt{2}}{2}g_{11}+\sqrt{2}g_{12})$ \\
$a_2^0\to \bar K^{*0}K^0$ & $\frac{\sqrt{2}}{2}g_{10}+(\frac{\sqrt{2}}{2}g_{11}-\sqrt{2}g_{12})$ \\
$a_2^-\to K^{*0}K^-$ & $g_{10}+(g_{11}-2g_{12})$ \\
$a_2^-\to K^{*-}K^0$ & $-g_{10}+(-g_{11}+2g_{12})$ \\
\hline
$f'_2\to K^{*+}K^-$ & $\frac{\sqrt{6}}{2}\cos\theta_T\,g_{10}+(-\frac{\sqrt{6}}{2}\cos\theta_T\,g_{11}-\sqrt{3}\sin\theta_T\,g_{11}+\sqrt{3}\sin\theta_T\,g_{12})$ \\
$f'_2\to K^{*-}K^+$ & $-\frac{\sqrt{6}}{2}\cos\theta_T\,g_{10}+(\frac{\sqrt{6}}{2}\cos\theta_T\,g_{11}+\sqrt{3}\sin\theta_T\,g_{11}-\sqrt{3}\sin\theta_T\,g_{12})$ \\
$f'_2\to K^{*0}\bar K^0$ & $\frac{\sqrt{6}}{2}\cos\theta_T\,g_{10}+(-\frac{\sqrt{6}}{2}\cos\theta_T\,g_{11}-\sqrt{3}\sin\theta_T\,g_{11}+\sqrt{3}\sin\theta_T\,g_{12})$ \\
$f'_2\to \bar K^{*0}K^0$ & $-\frac{\sqrt{6}}{2}\cos\theta_T\,g_{10}+(\frac{\sqrt{6}}{2}\cos\theta_T\,g_{11}+\sqrt{3}\sin\theta_T\,g_{11}-\sqrt{3}\sin\theta_T\,g_{12})$ \\
\hline
$f_2\to K^{*+}K^-$ & $\frac{\sqrt{6}}{2}\sin\theta_T\,g_{10}+(\sqrt{3}\cos\theta_T\,g_{11}-\frac{\sqrt{6}}{2}\sin\theta_T\,g_{11}-\sqrt{3}\cos\theta_T\,g_{12})$ \\
$f_2\to K^{*-}K^+$ & $-\frac{\sqrt{6}}{2}\sin\theta_T\,g_{10}+(-\sqrt{3}\cos\theta_T\,g_{11}+\frac{\sqrt{6}}{2}\sin\theta_T\,g_{11}+\sqrt{3}\cos\theta_T\,g_{12})$ \\
$f_2\to K^{*0}\bar K^0$ & $\frac{\sqrt{6}}{2}\sin\theta_T\,g_{10}+(\sqrt{3}\cos\theta_T\,g_{11}-\frac{\sqrt{6}}{2}\sin\theta_T\,g_{11}-\sqrt{3}\cos\theta_T\,g_{12})$ \\
$f_2\to \bar K^{*0}K^0$ & $-\frac{\sqrt{6}}{2}\sin\theta_T\,g_{10}+(-\sqrt{3}\cos\theta_T\,g_{11}+\frac{\sqrt{6}}{2}\sin\theta_T\,g_{11}+\sqrt{3}\cos\theta_T\,g_{12})$ \\
\hline
\end{tabular}
\end{center}\label{Tab:TtoPVamplitudes}
\end{table}

\begin{table}[htbp]
\centering
\renewcommand\arraystretch{0.96}
\caption{Branching ratios for the $T\to PV$ decays with $1\sigma$ uncertainties, in units of $10^{-2}$. The superscript E denotes an experimental input.}
\label{Tab:TtoPVBR}
\setlength{\tabcolsep}{2.5pt}
\begin{tabular}{@{}l@{\hspace{34pt}}c@{\hspace{12pt}}|@{\hspace{12pt}}l@{\hspace{34pt}}c@{}}
\hline\hline
Decay mode & Branching ratio &
Decay mode & Branching ratio\\
\hline
$\mathcal B(\bar K_2^{*0}\to\rho^+K^-)$ & $5.98\pm0.84$ & $\mathcal B(\bar K_2^{*0}\to\bar K^{*0}\eta)$ & $0.57\pm0.23$\\
$\mathcal B(\bar K_2^{*0}\to\rho^0\bar K^0)$ & $2.82\pm0.39$ & $\mathcal B(K_2^{*-}\to K^{*-}\eta)$ & $0.46\pm0.19$\\
$\mathcal B(K_2^{*-}\to\rho^0K^-)$ & $2.99\pm0.40$ & $\mathcal B(\bar K_2^{*0}\to\omega\bar K^0)$ & $2.85\pm0.88$\\
$\mathcal B(K_2^{*-}\to\rho^-\bar K^0)$ & $5.66\pm0.75$ & $\mathcal B(K_2^{*-}\to\omega K^-)$ & $3.00\pm0.93$\\
$\mathcal B(K_2^*\to\rho K)$ & $8.70\pm0.80\ExpMark$ & $\mathcal B(K_2^*\to\omega K)$ & $2.90\pm0.80\ExpMark$\\
$\mathcal B(\bar K_2^{*0}\to K^{*-}\pi^+)$ & $16.36\pm1.57$ & $\mathcal B(\bar K_2^{*0}\to\phi\bar K^0)$ & $0.074\pm0.056$\\
$\mathcal B(\bar K_2^{*0}\to\bar K^{*0}\pi^0)$ & $8.06\pm0.77$ & $\mathcal B(K_2^{*-}\to\phi K^-)$ & $0.047\pm0.034$\\
$\mathcal B(K_2^{*-}\to K^{*-}\pi^0)$ & $8.63\pm0.79$ & & \\
$\mathcal B(K_2^{*-}\to\bar K^{*0}\pi^-)$ & $16.56\pm1.51$ & & \\
$\mathcal B(K_2^*\to K^*\pi)$ & $24.70\pm1.50\ExpMark$ & & \\
\hline
$\mathcal B(a_2^0\to\rho^+\pi^-)$ & $34.97\pm1.53$ & $\mathcal B(a_2^0\to K^{*+}K^-)$ & $0.041\pm0.010$\\
$\mathcal B(a_2^0\to\rho^-\pi^+)$ & $34.95\pm1.47$ & $\mathcal B(a_2^0\to K^{*-}K^+)$ & $0.041\pm0.010$\\
$\mathcal B(a_2^-\to\rho^-\pi^0)$ & $35.40\pm1.51$ & $\mathcal B(a_2^0\to K^{*0}\bar K^0)$ & $0.034\pm0.0086$\\
$\mathcal B(a_2^-\to\rho^0\pi^-)$ & $34.92\pm1.47$ & $\mathcal B(a_2^0\to\bar K^{*0}K^0)$ & $0.034\pm0.0086$\\
$\mathcal B(a_2\to\rho\pi)$ & $70.10\pm2.70\ExpMark$ & $\mathcal B(a_2^-\to K^{*0}K^-)$ & $0.074\pm0.018$\\
& & $\mathcal B(a_2^-\to K^{*-}K^0)$ & $0.075\pm0.018$\\

\hline
$\mathcal B(f_2\to K^{*+}K^-)$ & $0.072\pm0.012$ & $\mathcal B(f'_2\to K^{*+}K^-)$ & $5.23\pm1.06$\\
$\mathcal B(f_2\to K^{*-}K^+)$ & $0.072\pm0.012$ & $\mathcal B(f'_2\to K^{*-}K^+)$ & $5.22\pm1.05$\\
$\mathcal B(f_2\to K^{*0}\bar K^0)$ & $0.064\pm0.011$ & $\mathcal B(f'_2\to K^{*0}\bar K^0)$ & $4.54\pm0.93$\\
$\mathcal B(f_2\to\bar K^{*0}K^0)$ & $0.064\pm0.011$ & $\mathcal B(f'_2\to\bar K^{*0}K^0)$ & $4.54\pm0.93$\\

\hline
\end{tabular}
\end{table}

\FloatBarrier

\section*{ACKNOWLEDGEMENTS}
The work was supported by the National Natural Science Foundation of China (Nos. 12365014 and 12665018).

\end{document}